\pdfoutput=1
\documentclass[twocolumn,superscriptaddress,aps,preprintnumbers,amsmath,amssymb,prd,nofootinbib]{revtex4}

\usepackage{amsmath}
\usepackage{amssymb}
\usepackage{amsfonts}
\usepackage{graphicx}
\usepackage{xcolor}
\usepackage{xfrac}
\usepackage{comment}
\usepackage{pifont}
\usepackage{physics}
\usepackage{fourier}
\usepackage{hyperref}
\usepackage{bm}
\usepackage{enumitem}

\definecolor{rossoferrari}{HTML}{D9073D}
\definecolor{mediumblue}{HTML}{0000CD}
\definecolor{forestgreen}{HTML}{228B22}
\definecolor{desy_blue}{HTML}{009EE2}
\definecolor{desy_orange}{HTML}{FD8800}
\definecolor{light_pink}{rgb}{1,0.4,0.4}
\definecolor{light_blue}{rgb}{0.284602,0.317763,0.963947}
\hypersetup{setpagesize=false,bookmarksnumbered=true,bookmarksopen=true,%
colorlinks=true,linkcolor=light_blue,urlcolor=rossoferrari,citecolor=rossoferrari,linktocpage=false}

\usepackage{slashed}

\def\bZ{{\mathbb Z}}

\def\SU{\mathrm{SU}}

\def\SO{\mathrm{SO}}
\def\Sp{\mathrm{Sp}}

\def\Spin{\mathrm{Spin}}

\def\beq#1\eeq{\begin{align}#1\end{align}}

\newcommand{\GeV}{\  {\rm GeV} }

\newcommand{\lmk}{\left(}  
\newcommand{\rmk}{\right)}

\newcommand{\eq}[1]{Eq.~(\ref{#1})}

\begin{document}


\preprint{TU-1153}


\title{Cosmic $F$- and $D$-strings from pure Yang--Mills theory}


\author{Masaki~Yamada}
\affiliation{Department of Physics, Tohoku University, Sendai, Miyagi 980-8578, Japan}
\affiliation{FRIS, Tohoku University, Sendai, Miyagi 980-8578, Japan}

\author{Kazuya~Yonekura}
\affiliation{Department of Physics, Tohoku University, Sendai, Miyagi 980-8578, Japan}

\date{\today}

\begin{abstract}
\noindent
We discuss the formation of cosmic strings or macroscopic color flux tubes after the deconfinement/confinement phase transition in the pure Yang--Mills theory. Based on holographic dual descriptions, these cosmic strings can be interpreted as fundamental (F-) strings or wrapped D-branes (which we call as D-strings) in the gravity side, depending on the structure of the gauge group. In fact, the reconnection probabilities of the F- and D-strings are suppressed by factors of $1/N^2$ and $e^{-c N}$, where $c = \mathcal{O}(1)$, in a large-$N$ limit, respectively. Supported by the picture of electric--magnetic duality, we discuss that color flux tubes form after the deconfinement/confinement phase transition, just like the formation of local cosmic strings after spontaneous symmetry breaking in the weak-U(1) gauge theory. We use an extended velocity-dependent one-scale model to describe the dynamics of the string network and calculate the gravitational wave signals from string loops. We also discuss the dependence on the size of produced string loops.
\end{abstract}


\maketitle


\section{Introduction}
The Universe is a unique system for the application of string theory, 
owing to the occurrence of high-energy phenomena that cannot be accessed by collider experiments. 
For example, the cosmologies of string landscapes~\cite{Bousso:2000xa,Susskind:2003kw}, string axions~\cite{Witten:1984dg,Svrcek:2006yi,Arvanitaki:2009fg}, swampland conjectures~\cite{Vafa:2005ui,Arkani-Hamed:2006emk,Garg:2018reu,Ooguri:2018wrx}, and brane inflationary scenarios~\cite{Dvali:1998pa} have been extensively reported in the literature. 
Among them, one of the most profound implications of string theory is the formation of macroscopic superstrings after a brane inflation~\cite{Dvali:2003zj,Copeland:2003bj}. 
Their dynamics is qualitatively different from that of field-theory cosmic strings, such that their reconnection probability is much smaller than unity~\cite{Jones:2003da,Jackson:2004zg}. 
These cosmic superstrings emit gravitational waves (GWs), which may be detected by GW observations~\cite{Vilenkin:1981bx,Vachaspati:1984gt}. This would be a unique way to prove the elemental components of string theory.

String theory also provides theoretical tools for understanding the strong dynamics of gauge theories. 
In this letter, we discuss that cosmic strings form after the deconfinement/confinement phase transition in the pure Yang--Mills (YM) theory, 
supported by the current understanding of theoretical physics. 
Based on a holographic dual description, these color flux tubes can be regarded as fundamental (F-) strings or wrapped D-branes (D-strings) in the gravity side, depending on the structure of the gauge group. This suggests that the properties of our cosmic strings should be similar to those of the superstrings produced after a D-brane inflation. 
Actually, the intercommutation probability of our cosmic strings is suppressed by a factor of $1/N^2$ or $e^{-cN}$, where $c = \mathcal{O}(1)$, in a large-$N$ limit, and can be much smaller than unity~\cite{Polchinski:1988cn,Jackson:2004zg,Hanany:2005bc}. 
Because the confinement is dual to the Higgsing in electric--magnetic duality, our cosmic strings should form in the phase transition. 
Therefore, the pure YM theory is the simplest model that creates F- and D-strings on the cosmological scale. We do not assume a brane inflationary scenario or the existence of extra dimensions, instead we just introduce the pure YM theory without quarks in a dark sector.

The statistical properties of the string network can be described by a velocity-dependent one-scale (VOS) model~\cite{Kibble:1984hp,Martins:1995tg,Martins:1996jp,Martins:2000cs}, 
which is supported by numerical simulations~\cite{Ringeval:2005kr,Blanco-Pillado:2011egf,Blanco-Pillado:2013qja,Blanco-Pillado:2017oxo,Blanco-Pillado:2017rnf}, 
and can be used to calculate GW signals~\cite{Caldwell:1991jj,DePies:2007bm,Sanidas:2012ee,Sousa:2013aaa,Sousa:2016ggw}. 
In this study, we adopt an extended VOS model proposed in Ref.~\cite{Avgoustidis:2005nv}, where the correlation length and interstring distance of long strings are treated separately to take into account the effect of a small reconnection (or intercommutation) probability. 
We numerically solve the extended VOS equation and calculate the GW spectrum emitted from cosmic F- and D-string loops. 

The details of the results discussed in the present letter are explained in the full paper~\cite{Yamada:2022imq}. In this letter, we also include the sensitivities of GW experiments on the size of string loops.

\section{F-strings in SU($N$) and Sp($N$)}
We consider the pure YM theory, where the gauge group is SU($N$), Sp($N$), or SO($N$). 
First, let us focus on the pure SU($N$) and Sp($N$) YM theories. 
We will comment on the case of SO($N$) subsequently. 

The pure YM theory is asymptotically free such that the gauge interaction becomes strong and is confined in low energies. 
We denote the dynamical scale at which the gauge coupling blows up as $\Lambda$. 
The dynamical scale is naturally small without any fine-tuning, owing to dimensional transmutation.

The deconfinement/confinement phase transition proceeds via the formation of color flux tubes, which can connect charged particles. In the pure YM theory, although there are no charged particles, one can still expect color flux tubes to form in the deconfinement/confinement phase transition. 
This is supported by the electric--magnetic duality demonstrated explicitly by some models~\cite{Seiberg:1994rs}, where 
the confinement is dual to the Higgsing 
and a color flux tube is dual to a vortex string in dual theory. 
The string tension, $\mu$, scales as $\mu \sim \Lambda^2$. 
From lattice simulations, the numerical factor is determined as $\mu \simeq 4 \Lambda^2$ 
for the SU($N$) gauge theory for a large $N$~\cite{Athenodorou:2021qvs}.

The dynamics of the string network depends on the reconnection or intercommutation probability of the strings. 
It can be estimated using the theory of a large-$N$ limit~\cite{tHooft:1973alw} (see \cite{Coleman:1985rnk} for a review).
The amplitude is proportional to the string coupling, $g_s$,
and it is estimated as $g_s \sim 1/N$ in the large-$N$ limit.
This result is further supported by holographic duality. 
In some confining gauge theories, 
there exist holographic dual descriptions that allow us to consider the theory in the gravity side (see e.g., Refs.~\cite{Witten:1998zw,Polchinski:2000uf,Klebanov:2000hb,Maldacena:2000yy,Vafa:2000wi}). 
A color flux tube in the SU($N$) YM theory is dual to an F- string in the gravity side. 
Because the 't Hooft coupling, $N g_s$, is fixed in the large-$N$ limit, the string coupling, $g_s$, scales as $N^{-1}$. 
The reconnection probability of an F-string is given by $g_s^2 \sim N^{-2}$. 
(See \cite{Jackson:2004zg} for more details of the computation of the amplitudes of the cosmic superstrings.)

As the strings are not stable in the existence of light quarks, we are interested in the case in which quarks are 
heavier than the dynamical scale, when they exist. 
If there is a heavy quark with mass $m$, 
the decay rate of a string via a quark/antiquark pair creation per unit volume is given by 
$\Gamma \sim \tilde{\mu}^2 \exp\left(  - \pi m^2/\mu \right)$, 
where $\tilde{\mu}$ is a typical mass scale related to the string and the particle~\cite{Vilenkin:1982hm}. 
The quark mass can be naturally of the same order as $\Lambda$ by a mechanism similar to that discussed in Refs.~\cite{Luty:2004ye,Ibe:2007wp,Yanagida:2010zz}, in which case the lifetime of a string can be shorter than the present age of the Universe. 
Such decaying cosmic strings have gained significant attention in recent works~\cite{Buchmuller:2021mbb,Dunsky:2021tih,Lazarides:2022jgr}. 
In this case, the quarks and anti-quarks should be diluted by inflation in order for sufficiently long strings to form at the deconfinement/confinement phase transition.

The stability of cosmic strings can be understood by a one-form symmetry, in a similar way to the stability of particles being ensured by an ordinary zero-form symmetry~\cite{Gaiotto:2014kfa} (see also Refs.~\cite{tHooft:1977nqb,tHooft:1979rtg,Witten:1985fp}). The one-form center symmetry of the pure YM theory for a simply connected gauge group $G$ can be determined by its center, which is a subgroup of $G$ whose elements commute with any element of $G$. The centers of SU($N$) and Sp($N$) are
\beq
\SU(N) \supset \bZ_N, \qquad 
\Sp(N) \supset \bZ_2.
\eeq
These centers $\bZ_N$ and $\bZ_2$ of the gauge groups lead to 
the corresponding one-form center symmetries, which we denote as $\bZ_N^{[1]}$ and $\bZ_2^{[1]}$, respectively. 

The $\bZ_N^{[1]}$ one-form symmetry implies that $N$ strings can join at a single vertex, called a baryon vertex, which should not be confused with a baryon particle. 
The network of such cosmic strings is similar to that of $\bZ_N$ cosmic strings considered in field-theory models~\cite{Vachaspati:1986cc}. 
In the case of the $\bZ_2^{[1]}$ one-form symmetry, 
a baryon vertex may or may not exist. If it exists, it just connects two strings. The network of such cosmic strings is sometimes called a necklace~\cite{Hindmarsh:1985xc,Berezinsky:1997td,Hindmarsh:2016dha}. The effect of a baryon vertex (or beads in the context of a necklace network) in this system can be neglected in the dynamics~\cite{Hindmarsh:2016dha}. 
The qualitative difference between SU($N$) and Sp($N$) is in their center symmetries; the latter one constructs a rather trivial string network, as we explained above. 
In this study, we neglect the effect of a baryon vertex, which is justified at least for the gauge group Sp($N$), including $\SU(2) = \Sp(1)$.
However, according to field-theory simulations, the effect of a baryon vertex of SU($N$) is not that significant even for $N=3$~\cite{Copeland:2005cy,Hindmarsh:2006qn,Urrestilla:2007yw}. 
The reconnection probability for Sp($N$) is still suppressed by $N^{-2}$.

\section{D-strings in SO($N$) }
Next, let us consider SO($N$) or $\Spin(N)$, which is a simply connected double cover of $\SO(N)$ ($=\Spin(N)/\bZ_2$). 
We specifically consider $\Spin(N)$ to explain the properties of the center symmetry; however, 
the cosmological implication of SO($N$) is the same as that of $\Spin(N)$. 
Its center is given by 
\beq
\Spin(N) \supset \left\{
\begin{array}{cl} \bZ_2 \times \bZ_2 & (N=4K) \\ \bZ_4 & (N=4K+2) \\ \bZ_2 & (N=2K+1) 
\end{array} \right. . 
\eeq
Thus, we can neglect the effect of a baryon vertex at least for the cases of $N = 4K$ and $2K+1$. 
Hereafter, we focus on $N = 2K+1$ for Spin($N$). 

A color flux tube of the Spin($N$) gauge theory is quite different from those of SU($N$) and Sp($N$). 
There are two different color flux tubes in Spin($2K+1$): one type is created by a fundamental ($N$-dimensional) representation of the gauge group, 
and the other is created by a spinor representation. 
A color flux tube associated with the fundamental representation is metastable in Spin($2K+1$) because 
a colored baryon exists in the fundamental representation. 
Actually, one can write the following operator: 
\beq
B_{i, \mu_1 \cdots \mu_{2K}}  = \epsilon_{i i_1 \cdots i_{2K}} F^{i_1i_2}_{\mu_1\mu_2} \cdots F^{i_{2K-1}i_{2K}}_{\mu_{2K-1} \mu_{2K}}, 
\eeq
where $F^{ij}_{\mu\nu}$ is the gauge field strength with color indices $i,j$ and spacetime indices $\mu,\nu$.
This operator can create a baryon with color index $i$. 
As we discussed above, a color flux tube associated with the fundamental representation can decay via baryon pair production. 
The estimated mass of a baryon is of order $N \Lambda$~\cite{Witten:1979kh} and a string can be of long duration for a large $N$. 
However, baryons and anti-baryons also form at the phase transition and it is expected that all F-strings end on them at the phase transition. Therefore long F-strings cannot form in this gauge theory.

A stable color flux tube is also created by the spinor representation in Spin($N$). 
We expect that its tension is of order $N \Lambda^2$, as we explained in Ref.~\cite{Yamada:2022imq}. 
This is actually confirmed by a holographic dual description~\cite{Witten:1998xy}, 
where a color flux tube is dual to a wrapped D-brane with tension of order $g_s^{-1} \sim N$ in the gravity side. 
We call this type of cosmic string as a D-string.

Moreover, the reconnection probability of a D-string is exponentially suppressed as 
$P \sim e^{-cN}$ with $c = \mathcal{O}(1)$ for a relative velocity of $v = \mathcal{O}(1)$. 
This is similar to the case of a D1-brane formed after a brane inflation~\cite{Jones:2002cv,Sarangi:2002yt,Dvali:2002fi,Jones:2003da,Pogosian:2003mz,Dvali:2003zj,Copeland:2003bj}, where the coefficient, $c$, depends on the relative velocities and relative angles of the strings~\cite{Jackson:2004zg}. 
The probability may become $\mathcal{O}(1)$ for $v \sim \mathcal{O}(1/N)$. We explain these points in more detail in \cite{Yamada:2022imq}.

In summary, the string tension, $\mu$, and the intercommutation (or reconnection) probability, $P$, scale as
\beq
\label{eq:muP}
\begin{array}{c|c|c}
&\text{F-string}& \text{D-string} \\ \hline
\mu~~ &\Lambda^2& N\Lambda^2  \\ \hline
P~~ &N^{-2} & \exp(-c N)  
\end{array}
\eeq
in the large-$N$ limit, at least for $v = \mathcal{O}(1)$. 
SU($N$) and Sp($N$) contain 
only F-strings. 
Spin($2K+1$), 
includes both strings; however, long F-strings are not expected to form.

In the remainder of this letter, we do not assume a specific gauge group, 
but consider a single type of cosmic string with parameters $\mu$ and $P$.

\section{Extended VOS model}
Herein, we consider the dynamics of cosmic F- and D-strings. 
We calculate the evolution of the cosmic strings based on Ref.~\cite{Avgoustidis:2005nv}, in which the VOS model was extended to take into account the small intercommutation probability. 
The referred study also validated that the extended model was consistent with numerical simulations with a small $P$.

The consequence of a small intercommutation probability is nontrivial for the evolution of the string network. The cosmic strings have wiggly structures on a small scale, because they have no efficient energy-loss mechanism to make them smoother. 
The small wiggles in the strings move as fast as $1/\sqrt{2}$, whereas the relative velocity between long strings is not that high according to numerical simulations~\cite{Martins:2005es,Avgoustidis:2005nv}. 
This suggests that many intersections of small wiggles occur when two (wiggly) long strings collide. 
The number of intersections of small wiggles per unit collision event for long strings is estimated as $N_{\rm scat} \sim 10$~\cite{Avgoustidis:2005nv}. 
Thus, the effective intercommutation probability of wiggly long strings is given by 
\beq
 P_{\rm eff} = 1 - \lmk 1- P \rmk^{N_{\rm scat}}, 
 \label{eq:Peff}
\eeq
which approximates as $P_{\rm eff} \simeq N_{\rm scat} P \sim 10 P$ for a small $P$.

In the case of a small $P_{\rm eff}$, 
the correlation length, $\xi$, and the interstring distance of long strings, $L$, should be treated separately. 
The correlation length represents the distance beyond which the string directions are not correlated. 
It is decreased mainly by self-reconnection of the wiggly structures~\cite{Sakellariadou:2004wq}, 
which is described by left and right movers of string perturbations. 
These movers collide many times because of their periodic motion, and eventually reconnect. 
Thus, we expect that 
the rate of self-reconnection of a long string is not reduced even for $P_{\rm eff} \ll 1$, 
and hence, the correlation length is expected to be of the order of the Hubble length, namely, $\xi = c_\xi t$, where $c_\xi =\mathcal{O}(1)$. 
Ref.~\cite{Avgoustidis:2005nv} assumed $c_\xi = 1$ for simplicity. 
However, we take $c_\xi$ such that the standard results of the VOS model can be reproduced in the scaling regime for the case of $P_{\rm eff} =1$. This can be realized by $c_\xi = 0.27$ in a radiation- dominated era (RD) and by $c_\xi = 0.62$ in a matter-dominated era (MD). 
We interpolate these values in the period between them. 

The interstring distance of long strings, $L$, is defined by their energy density, $\rho_{\infty}$ by $\rho_{\infty} = \mu L^2$. 
Because the rate of collisions between long strings is reduced by $P_{\rm eff} \ll 1$, 
the number of long strings within the Hubble horizon can be larger than unity. 
This implies that $L$ can be shorter than the Hubble length. Thus, we need to treat $L$ and $\xi$ separately. 
The energy density of long strings decreases via the loop production function as follows: $(d\rho/dt)_{\rm loop} = - P_{\rm eff} (\tilde{c} \mu \xi) n (n \xi^3 \bar{v}/\xi)$. 
In this expression, 
$n$ ($=\rho_\infty / (\mu \xi) = 1/ (L^2 \xi)$) is the number density of long strings 
and the last parenthesis comes from 
the number of intersections for a given string per unit time ($\sim (n \xi^3 (\bar{v}/\xi)$). 
The coefficient, $\tilde{c}$, is determined by 
numerical simulations, such as 
$\tilde{c} = 0.23 \pm 0.04$ in both the RD and MD~\cite{Martins:2003vd} (see also Refs.~\cite{Bennett:1989yp,Allen:1990tv,Martins:1996jp,Martins:2000cs}). 
The evolution equations of $L$ and averaged velocity $\bar{v}$ in the extended VOS are then given by 
\beq
\label{eq:VTS1}
 &2 \frac{dL}{dt} = 2 H L \lmk 1 + \bar{v}^2 \rmk 
 + P_{\rm eff} \tilde{c} \bar{v} \lmk \frac{\xi}{L} \rmk, 
 \\
\label{eq:VTS2}
 &\frac{d\bar{v}}{dt} = \lmk 1 - \bar{v}^2 \rmk \lmk \frac{k(\bar{v})}{\xi} - 2 H \bar{v} \rmk, 
\eeq
where 
the momentum parameter, $k(\bar{v})$, is given by 
$k(\bar{v}) = (2\sqrt{2}/\pi) (1-8 \bar{v}^6)/(1+8 \bar{v}^6)$ 
in the relativistic regime~\cite{Martins:2000cs}. 
In the scaling regime, 
we have $L_{\rm asym} \propto \sqrt{P_{\rm eff}}$, which 
is consistent with the analytic argument and numerical simulations in the flat spacetime~\cite{Sakellariadou:1990nd, Sakellariadou:2004wq}. 
Moreover, the extended VOS model can explain the numerical simulations in both the RD and MD~\cite{Avgoustidis:2005nv}. 

\section{Gravitational wave signals}
GWs are mainly emitted from string loops, 
whose number density can be calculated from $L(t)$ and $\bar{v}(t)$ as follows:
$n_{\rm loop}(l,t) dl
= \rho_{\rm loop} (l_i, t) /(\mu l_i) dl_i$, 
where $l = l_i - \Gamma G \mu (t-t_i)$ 
with $\Gamma \approx 50$ being a numerical factor~\cite{Vachaspati:1984gt,Burden:1985md,Garfinkle:1987yw,Blanco-Pillado:2017oxo}
and 
\beq
  \rho_{\rm loop} (l_i,t) = 
P_{\rm eff} \mu l_i  \int_{t_i}^{t} dt'
  \lmk \frac{a(t')}{a(t)} \rmk^3 
   \frac{\bar{v}(t') }{L^4(t')} f (l_i,t'). 
\eeq
Here, $f(l_i,t)$ is a scale-invariant loop production function, which is the number of produced loops per unit length for each intercommutation event. 
In our convention, it is related to $\tilde{c}$ as follows:
$\int_0^{\infty}dl_i \, f(l_i,t) l_i  = \tilde{c} \xi$ 
and $f(l_i,t) = t^{-1} f(l_i/t)$. 
In previous studies, it is conventionally assumed that $f(l_i/t)$ is monotonic, expressed as follows:
\beq
f(x) 
= \frac{\mathcal{F}}{f_r} \frac{\tilde{c} c_\xi }{\alpha} \delta (x - \alpha),
\eeq
where 
$f_r = \sqrt{2}$ comes from the redshift of the string loops~\cite{Vilenkin:2000jqa}. 
Moreover, $\mathcal{F} = 0.1$ is introduced to incorporate the finite width effect of the string loop spectrum~\cite{Sanidas:2012ee,Blanco-Pillado:2013qja}. 
Because the correlation length or a typical curvature of each long string is of order $\xi$, we expect that the loop size is proportional to $\xi$, rather than $L$, and is independent of $P_{\rm eff}$. 
The value of $\alpha$ is under debate in the literature, even in the case of $P_{\rm eff} = 1$. 
In this letter, we mainly adopt $\alpha =0.1$, which is confirmed by Nambu--Goto simulations with $P_{\rm eff} =1$~\cite{Blanco-Pillado:2013qja}. We also consider the case with $\alpha \ll 0.1$ to show its effect on the GW spectrum in Sec.~\ref{sec:alpha}.

The GW spectrum at present is calculated from  
\beq
 \frac{d \rho_{\rm GW}}{d \ln f}
 = 
 G\mu^2 \sum_{n=1}^\infty P_n
 \frac{2n}{f} 
 \int_0^{z_i} \frac{dz}{H(z) (1+z)^6} 
 n_{\rm loop} \lmk \frac{2n}{(1+z)f}, t'(z) \rmk,
 \label{eq:Omegagw}
\eeq
where $z$ is the redshift 
and 
the averaged power spectra is given by 
$P_n = \Gamma n^{-q}/\xi(q)$, 
where $\xi(q)$ is the zeta function. 
We take $q = 4/3$, assuming that GWs are dominantly produced by cusps.


\begin{figure}
\includegraphics[width=0.475\textwidth]{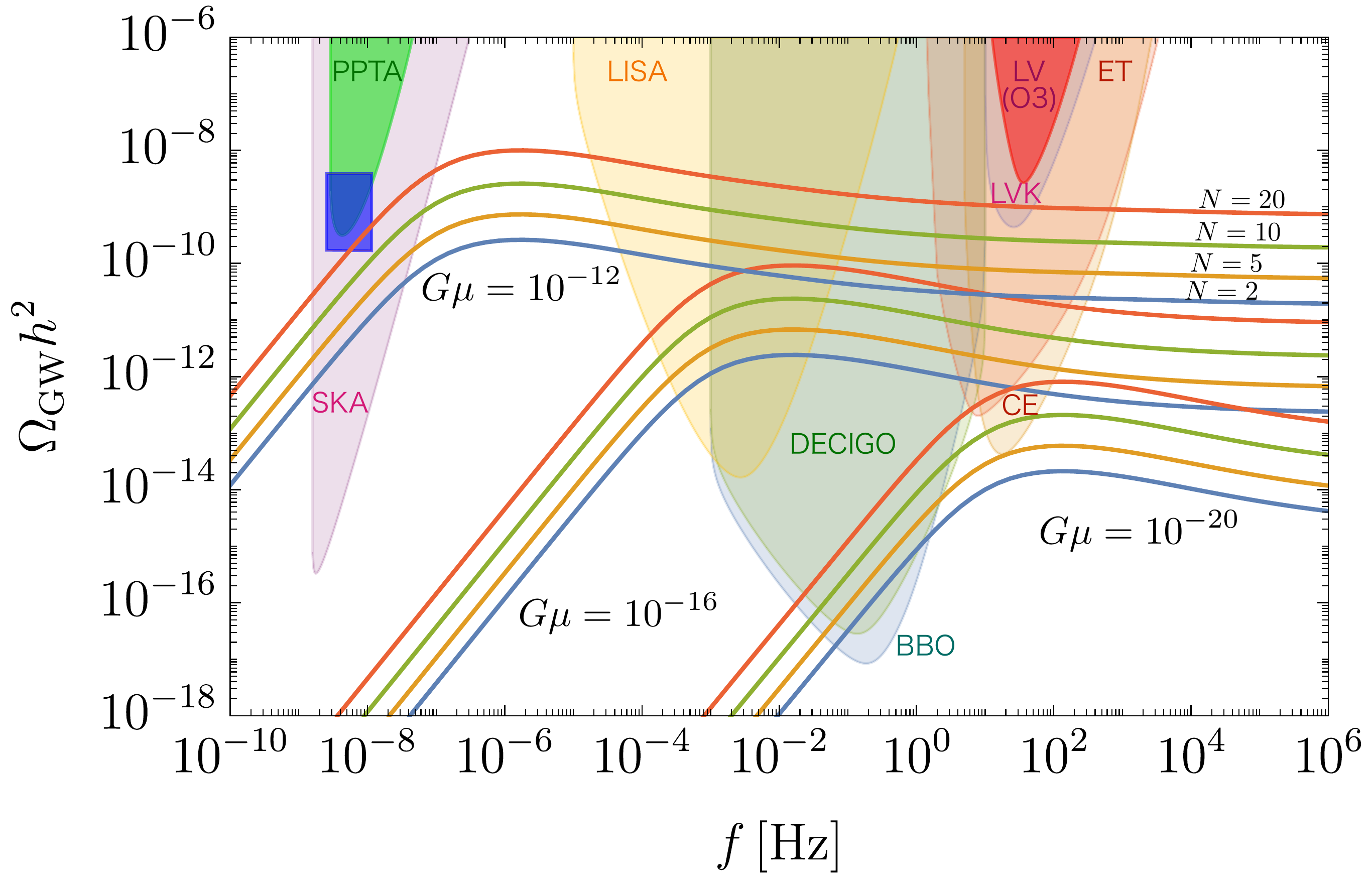}
\caption{GW spectrum emitted from cosmic string loops. 
We take $G\mu = 10^{-12}$, $10^{-16}$, and $10^{-20}$ (from top to bottom) with 
$N=2$ (blue), $5$ (yellow), $10$ (green), and $20$ (red). For the sensitivity curves and current constraint, see the main text. 
}
\label{fig:1}
\end{figure}



\begin{figure}
\includegraphics[width=0.475\textwidth]{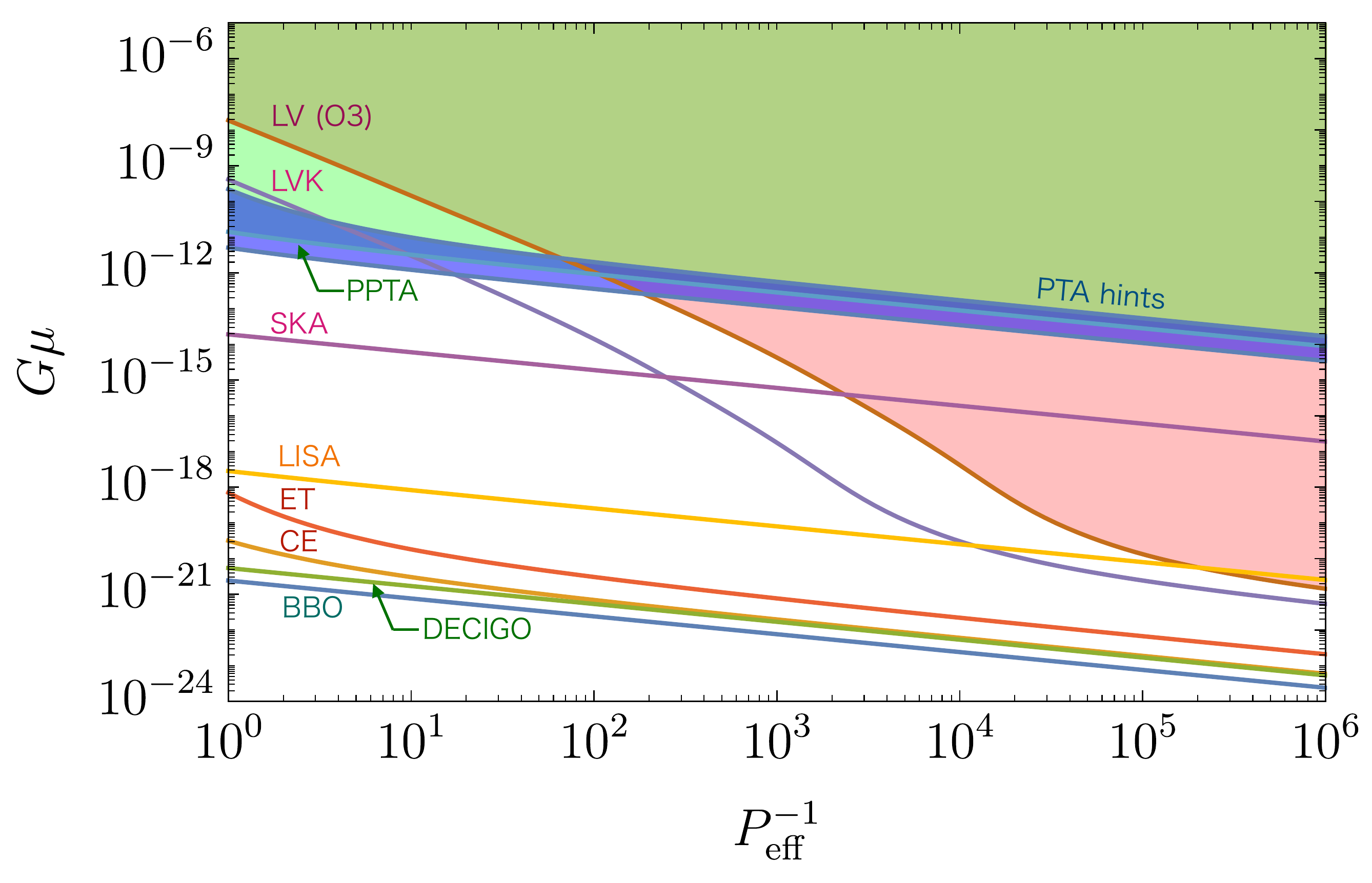}
\caption{Sensitivity curves for ongoing and planned GW experiments in the $G\mu$-$P_{\rm eff}^{-1}$ plane. 
The green and red shaded regions are excluded by the current experiments. The blue shaded region is favored to explain the PTA hints. 
}
\label{fig:2}
\end{figure}


\subsection{Results}

We numerically solve the extended VOS equations and calculate the GW spectrum, 
where we take 
$\Omega_\Lambda = 0.685$ and 
$\Omega_m = 0.315$~\cite{Planck:2018vyg} as the cosmological parameters. 
We take into account the entropy production by decoupling the relativistic degrees of freedom in the Standard--Model sector. 
The resulting GW spectra are shown in Fig.~\ref{fig:1} 
for the cases of $G\mu = 10^{-12}$, $10^{-16}$, and $10^{-20}$ with 
$N=2,5,10$, and $20$. 
We take $\alpha = 0.1$ and $N_{\rm scat} = 10$. 
The peak amplitude of the GWs and the peak frequency are approximately given by 
\beq
\label{eq:resultGW}
 &\lmk \Omega_{\rm GW} h^2 \rmk^{(\rm peak)} \simeq 2.5 \times 10^{-10} \times P_{\rm eff}^{-1} \lmk \frac{G\mu}{10^{-12}} \rmk^{1/2} 
 \lmk \frac{\alpha}{0.1} \rmk^{1/2}, 
 \\
\label{eq:resultfreq}
 &f^{(\rm peak)} \simeq 1.9 \times 10^{-6} \times \lmk \frac{G\mu}{10^{-12}} \rmk^{-1}, 
\eeq
for $\alpha \gg \Gamma G \mu$.
Because the dependence of $P_{\rm eff}^{-1}$ and $G\mu$ are not degenerate, we can determine both by observing the spectra around the peaks. 

We follow Ref.~\cite{Schmitz:2020syl} to plot 
the power-law-integrated sensitivity curves for ongoing and planned GW experiments: 
SKA~\cite{Janssen:2014dka},
LISA~\cite{LISA:2017pwj},
DECIGO~\cite{Kawamura:2011zz,Kawamura:2020pcg},
BBO~\cite{Harry:2006fi},
Einstein Telescope (ET)~\cite{Punturo:2010zz,Maggiore:2019uih},
Cosmic Explorer (CE)~\cite{Reitze:2019iox},
and aLIGO+aVirgo+KAGRA (LVK)~\cite{Somiya:2011np,KAGRA:2020cvd}.
The current constraints from the Parkes pulsar timing array (PPTA)~\cite{Shannon:2015ect} 
and aLIGO/aVirgo's third observing run (LV(O3))~\cite{KAGRA:2021kbb} 
are shown as green and red shaded regions, respectively. 
The blue box highlights the potential signals of pulsar timing array (PTA) experiments: NANOGrav~\cite{NANOGrav:2020bcs} and the PPTA~\cite{Goncharov:2021oub}. 
See Refs.~\cite{Ellis:2020ena,Blasi:2020mfx,Blanco-Pillado:2021ygr} for 
other cosmic string models to explain the PTA hints.

Figure~\ref{fig:2} shows the sensitivity curves in the $G\mu$-$P_{\rm eff}^{-1}$ plane. 
The red and green shaded regions are excluded by the current experiments. 
The blue shaded region is favored to explain the PTA hints. 
Interestingly, we can predict the signals in LVK consistently with the PTA experiments for $P_{\rm eff}^{-1} \gtrsim 10$. 
This is because the peak amplitude can be enhanced without changing the peak frequency for a large $P_{\rm eff}^{-1}$.

We note that $\mu$ and $P_{\rm eff}$ can be interpreted as physical quantities $\Lambda$ and $N$ from Eqs.~(\ref{eq:muP}) and (\ref{eq:Peff}), with $\mathcal{O}(1)$ uncertainties. 
In particular, for the case of F-strings, 
GW signals can be observed for $\Lambda \gtrsim 10^8 \GeV$ and $N = \mathcal{O}(1\,\text{-}\,10)$. 
For the case of D-strings, 
one should be careful 
about the exponentially suppressed intercommutation probability 
because it may become $\mathcal{O}(1)$ for $v \sim \mathcal{O}(1/N)$. 
Collisions with such a low relative velocity may not be negligible in the random motion of the cosmic string network. 
We do not further discuss this issue in this letter and leave it for a future work. 

\subsection{$\alpha$-dependence}
\label{sec:alpha}
\begin{figure}
\includegraphics[width=0.475\textwidth]{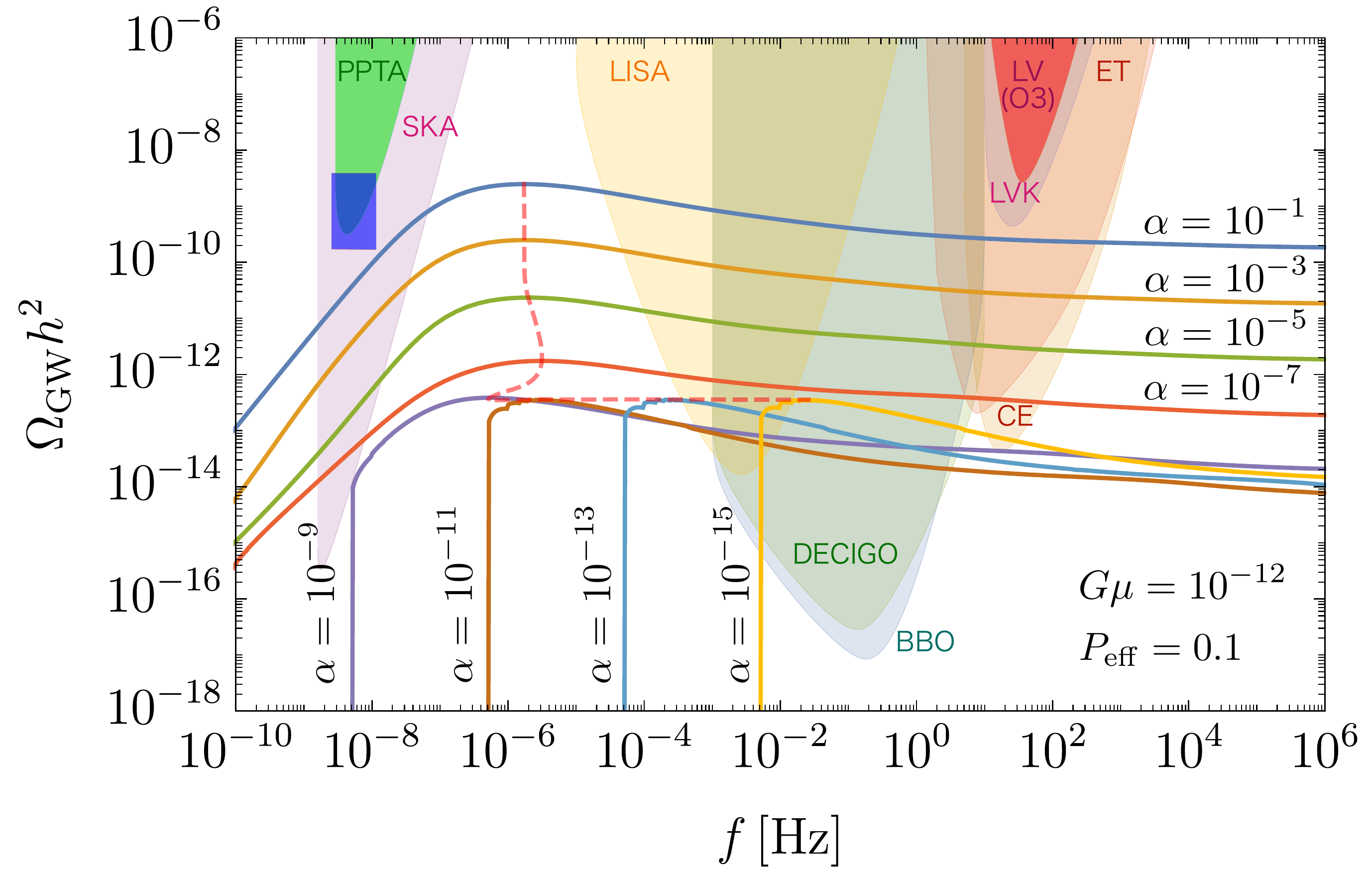}
\caption{Same as Fig.~\ref{fig:1} but with $G \mu = 10^{-12}$ and $P_{\rm eff} = 0.1$ 
for the case of $\alpha = 10^{-1}$, $10^{-3}$, $10^{-5}$, $10^{-7}$, $10^{-9}$, $10^{-11}$, $10^{-13}$, and $10^{-15}$ from top left to bottom right.
The red dashed curve represents the contour of the maximum of the spectrum, i.e., ($f^{(\rm peak)}, \lmk \Omega_{\rm GW} h^2 \rmk^{(\rm peak)}$), for $\alpha$ ranging from $10^{-1}$ to $10^{-15}$. 
}
\label{fig:3}
\end{figure}

Finally, we show the dependence of GW spectrum on the size of string loops $\alpha$. 
Figure~\ref{fig:3} shows the GW spectrum for the case of $G \mu = 10^{-12}$ 
and $P_{\rm eff} = 0.1$ 
with $\alpha = 10^{-1}$, $10^{-3}$, $10^{-5}$, $10^{-7}$, $10^{-9}$, $10^{-11}$, $10^{-13}$, and $10^{-15}$ from top left to bottom right. 
The red dashed curve represents the contour of the maximum of the spectrum, i.e., ($f^{(\rm peak)}, \lmk \Omega_{\rm GW} h^2 \rmk^{(\rm peak)}$), for $\alpha$ ranging from $10^{-1}$ to $10^{-15}$. 
As in the standard scenario of VOS model, 
the peak amplitude and frequency of GW spectrum depends on $\alpha$ such as $\lmk \Omega_{\rm GW} h^2 \rmk^{(\rm peak)} \propto \alpha^{1/2}$ and $f^{(\rm peak)} \propto \alpha^0$ for $\alpha \gg \Gamma G \mu$ and $\lmk \Omega_{\rm GW} h^2 \rmk^{(\rm peak)} \propto \alpha^0$ and $f^{(\rm peak)} \propto \alpha^{-1}$ for $\alpha \ll \Gamma G\mu$ (see, e.g., Ref.~\cite{Auclair:2019wcv}). 
The numerical factors are determined by our numerical calculations and are 
given by 
Eqs.~(\ref{eq:resultGW}) and (\ref{eq:resultfreq}) 
for $\alpha \gg \Gamma G \mu$ and 
\beq
\label{eq:resultGW2}
 &\lmk \Omega_{\rm GW} h^2 \rmk^{(\rm peak)} \simeq 3.8 \times 10^{-14} \times P_{\rm eff}^{-1} \lmk \frac{G\mu}{10^{-12}} \rmk, 
 \\
\label{eq:resultfreq2}
 &f^{(\rm peak)} \simeq 5.2 \times 10^{-7} \times \lmk \frac{G\mu}{10^{-12}} \rmk^{-1} \lmk \frac{\alpha }{ \Gamma G \mu} \rmk^{-1}, 
\eeq
for $\alpha \ll \Gamma G \mu$.


As $\alpha$ decreases from a relatively large value and approaches to $\Gamma G \mu$, the peak frequency slightly increases, decreases, and then increases again. 
This is in contrast to the case with the standard scenario of VOS model, 
where the peak frequency slightly decreases and then increases~\cite{Sanidas:2012ee}. 
This is because 
the last term in \eq{eq:VTS1} is modified in the extended VOS model, 
which then leads to a different behavior on the $\alpha$ dependence around $\alpha \sim \Gamma G\mu$. 
However, the difference is not significant or important for our purpose. 

For the case of $\alpha \ll \Gamma G \mu$, 
the spectrum has a lower cutoff on the frequency 
and 
is oscillating around the cutoff. 
This is because 
the contribution to the summation of \eq{eq:Omegagw} 
comes only from $n = 1$ just above the cutoff 
and higher modes come into the contribution as the frequency increases. 
Since $n$ must be integer and each modes contributes with a similar order of magnitude, 
the spectrum increases discretely as the frequency increases. 

\begin{figure}
\includegraphics[width=0.475\textwidth]{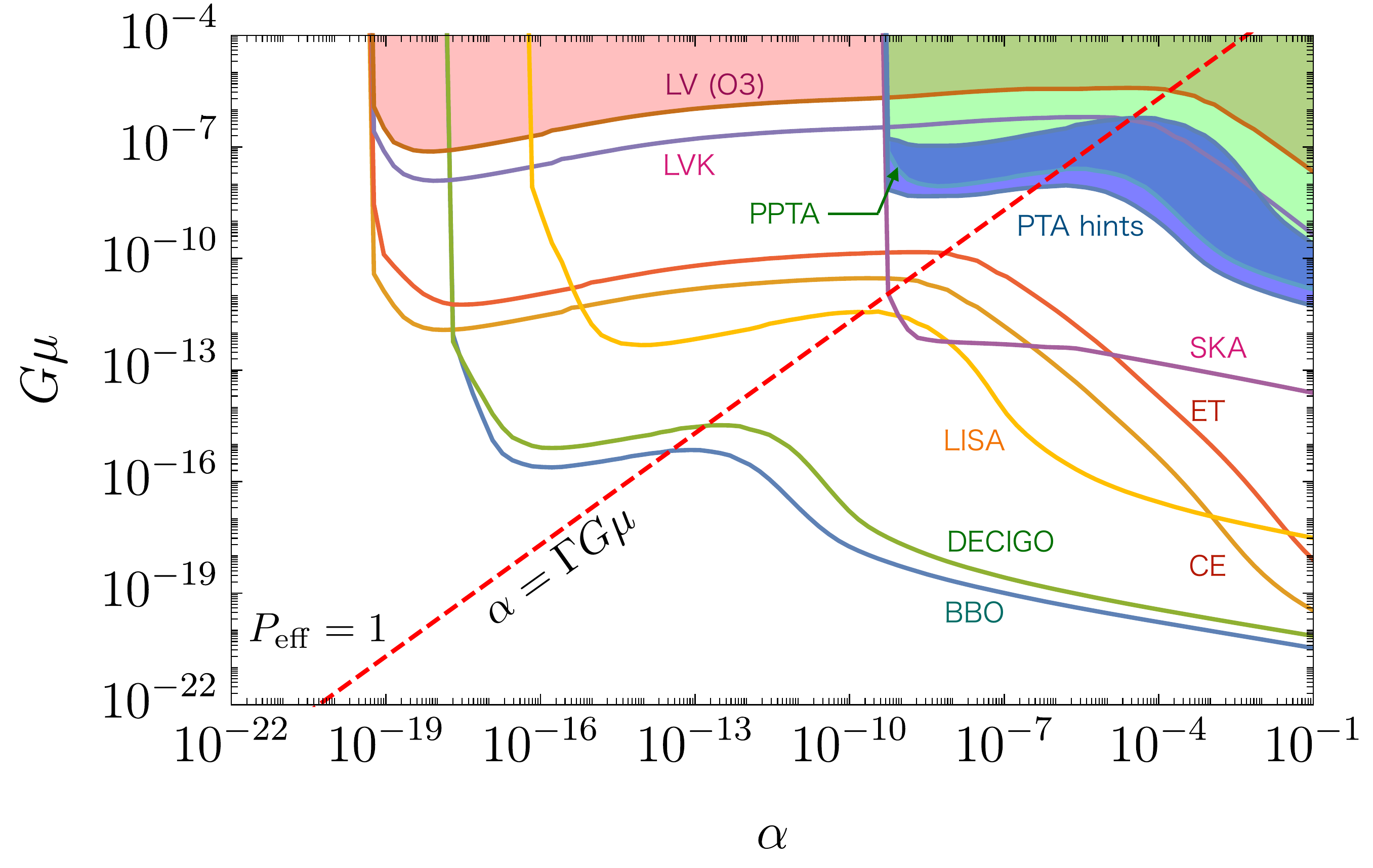}
\\
\includegraphics[width=0.475\textwidth]{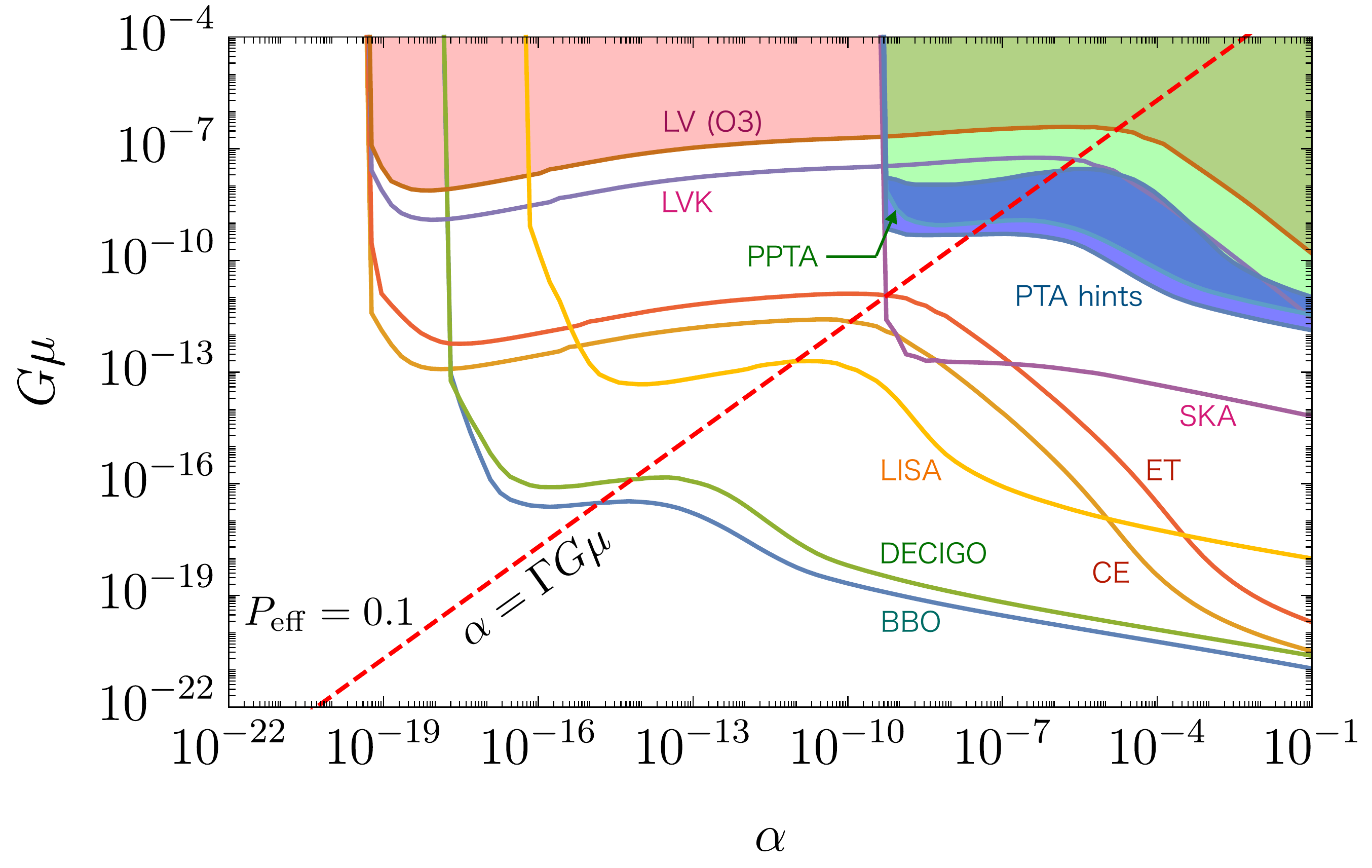}
\\
\includegraphics[width=0.475\textwidth]{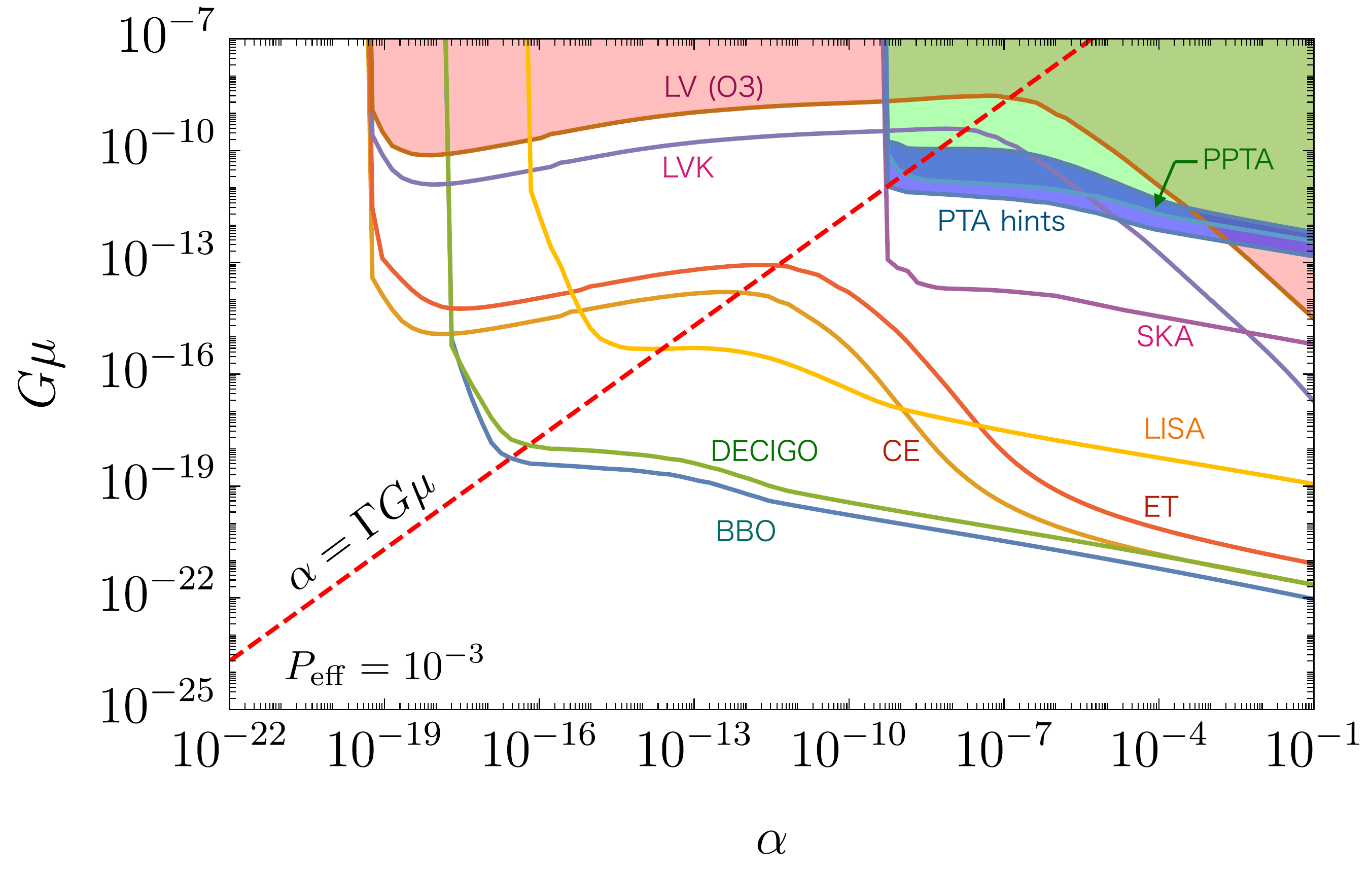}
\\
\includegraphics[width=0.475\textwidth]{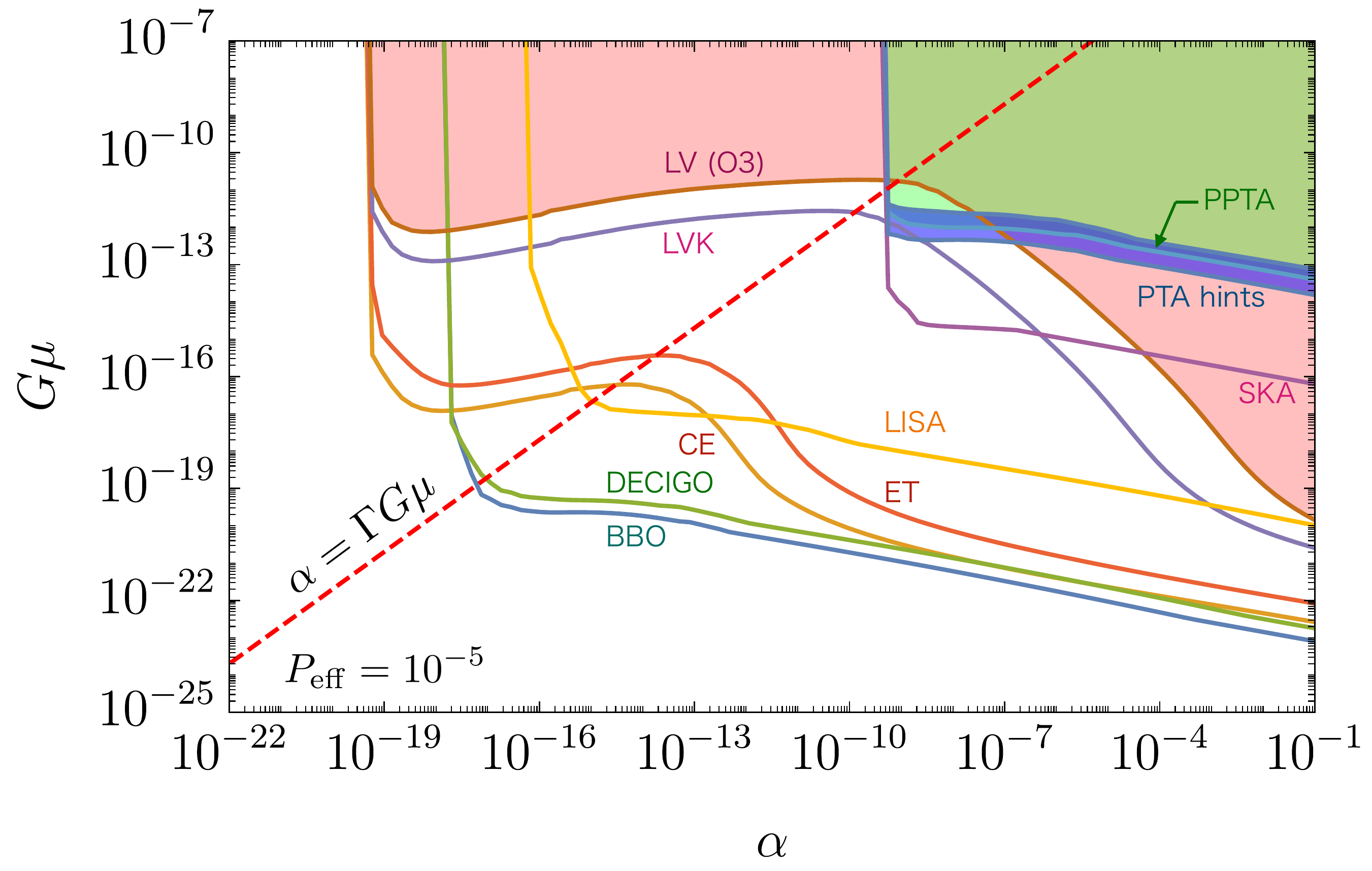}
\caption{Same as Fig.~\ref{fig:2} but with $P_{\rm eff} = 1$, $10^{-1}$, $10^{-3}$, and $10^{-5}$ (from top to bottom)) in the $G\mu$-$\alpha$ plane. 
The dashed line represents $\alpha = \Gamma G \mu$. 
}
\label{fig:4}
\end{figure}

Figure~\ref{fig:4} shows the sensitivity curves in the $G\mu$-$\alpha$ plane. 
We take $P_{\rm eff} = 1$, $10^{-1}$, $10^{-3}$, and $10^{-5}$ from top to bottom. 
The dashed line represents the value at $\alpha = \Gamma G\mu$. 
Because of the different parameter dependence of the GW spectrum for $\alpha \gg \Gamma G \mu$ and $\alpha \ll \Gamma G \mu$, 
the sensitivity curves have different behavior for those limits. 
If we adopt $\alpha \sim \Gamma G \mu$ rather than $\alpha = 0.1$, 
the sensitivity on $G \mu$ by LISA is weakened by a factor of $\mathcal{O}(10^{-6})$ 
for the case of $P_{\rm eff} =\mathcal{O}(0.1)$, 
whereas it is $\mathcal{O}(10^{-3})$ for the case of $P_{\rm eff} =\mathcal{O}(10^{-5})$.


\section{Discussion and conclusions}
We have discussed the formation and properties of macroscopic color flux tubes or cosmic strings in pure YM theories, such as SU($N$), Sp($N$), and SO($N$), and calculated their GW signals. 
These cosmic strings have a small intercommutation probability for a large $N$, just like cosmic superstrings; however, we do not assume a brane inflationary scenario or extra dimensions. 
The resulting GW signals can be observed by some ongoing and planned GW experiments if the dynamical scale is $\mathcal{O}(10^{8\,\text{-}\,13})\GeV$ for $N = \mathcal{O}(1\,\text{-}\,10)$ and $\alpha = 0.1$. 
We also clarify how the GW spectrum changes for a smaller loop size $\alpha$.

We have considered the case with a single type of cosmic strings without a baryon vertex to calculate GW signals. 
The $\SO(N)$ gauge theory may contain F-strings as well as D-strings. The tensions and intercommutation probabilities of these cosmic strings have different dependence on $N$ for a large $N$. The dynamics of the network can be described by introducing a VOS equation for each string with transition terms~\cite{Avgoustidis:2007aa}. 
The effect of a baryon vertex, such as in the case of $\SU(N)$ with $N \ge 3$, is not fully understood particularly for a large $N$. We expect that baryon vertices form after the phase transition and their number density should also have a scaling behavior. 
We leave these topics for a future work, as detailed numerical simulations may be necessary.

In previous studies, the YM theory with the deconfinement/confinement phase transition has been extensively considered for the formation of glueballs~\cite{Morningstar:1999rf,Lucini:2010nv,Curtin:2022tou}, which is a candidate for dark matter~\cite{Faraggi:2000pv,Feng:2011ik,Boddy:2014yra,Boddy:2014qxa,Soni:2016gzf,Kribs:2016cew,Forestell:2016qhc,Soni:2017nlm,Forestell:2017wov,Jo:2020ggs}. 
Our results suggest that such a scenario inevitably results in the formation of F- and/or D-strings and predicts GW signals from string loops. 
However, these glueballs may decay before the big-bang nucleosynthesis epoch in the case of $\Lambda \gtrsim 10^8 \GeV$~\cite{Juknevich:2009ji,Juknevich:2009gg,Halverson:2016nfq,Asadi:2022vkc}, which is the parameter of our interest.

We note that we only assume that the phase transition occurs after inflation. In particular, the phase transition can occur during an inflation—oscillation-dominated era. Even in this case, our result does not change because the GWs are emitted from string loops in the scaling regime of the RD. 
For the same reason, our calculation does not change even if the temperature of the gauge sector is different from that of the Standard--Model sector. 

\

\begin{acknowledgments}
The authors would like to thank Yuya Tanizaki for intensive lectures at Tohoku University which inspired this work.
MY thanks Alexander Vilenkin for valuable comments on metastable cosmic strings. 
MY also thanks Ken D. Olum and Jose J. Blanco-Pillado for useful discussions. 
MY was supported by MEXT Leading Initiative for Excellent Young Researchers, and by JSPS KAKENHI Grant No.\ 20H0585 and 21K13910. The work of KY is supported in part by JST FOREST Program (Grant Number JPMJFR2030, Japan), 
MEXT-JSPS Grant-in-Aid for Transformative Research Areas (A) ”Extreme Universe” (No. 21H05188),
and JSPS KAKENHI (17K14265).
\end{acknowledgments}


\bibliographystyle{JHEP}
\bibliography{ref}


\end{document}